\documentstyle[psfig,12pt]{article}
\title{Possible Tomography of the Sun's Magnetic Field with Solar Neutrinos}
\author{\bf Jo\~{a}o Pulido \\
Centro de F\'{\i}sica das Interac\c{c}\~{o}es Fundamentais \\
Instituto Superior T\'{e}cnico \\
Av. Rovisco Pais, 1096 Lisboa Codex, Portugal}

\newcommand{\be}{\begin{equation}}
\newcommand{\ee}{\end{equation}}
\newcommand{\bea}{\begin{eqnarray}}
\newcommand{\eea}{\end{eqnarray}}
\begin{document}
\maketitle

\begin{abstract}
The data from solar neutrino experiments together with standard solar model 
predictions are used in order to derive the possible profile of the magnetic 
field inside the Sun, assuming the existence of a sizeable neutrino magnetic 
moment and the resonant spin flavour mechanism. The procedure is based on the 
relationship between resonance location and the energy dependent neutrino
suppression, so that a large neutrino suppression at a given energy is taken
to be connected to a large magnetic field in a given region of the Sun. In 
this way it is found that the solar field must undergo a very sharp increase 
by a factor of at least 6 - 7 over a distance no longer than 7 - 10\% of the 
solar radius, decreasing gradually towards the surface. The range in which 
this sharp increase occurs is likely to be the bottom of the convective zone. 
There are also indications in favour of the downward slope being stronger at 
the start and more moderate on approaching the solar surface. Typical ranges 
for the magnetic moment are from a few times $10^{-13}\mu_B$ to its laboratory
upper bounds while the mass square difference between neutrino flavours is
${\Delta}^2m_{21}\simeq(0.6-1.9)\times10^{-8}eV^2$.
\end{abstract}

\newpage
{\bf \large 1. Introduction}

\vspace { 6mm}

It is usually argued that the solar neutrinos which were first observed in
1968 \cite {Davis} have not provided so far, as opposed to previous 
expectations, any particular knowledge of the physics inside the Sun. In fact 
the solar interior is opaque and detailed information about its structure has
turned out possible to obtain, not via direct probes, but indirectly, via
helioseismology \cite {IDFR}.

Neutrinos however hold the potential of real probes of the solar interior,
since they can travel across the Sun practically unperturbed. A neutrino 
produced near the centre of the Sun typically has a collision probability of 
the order of $10^{-6} - 10^{-7}$ while travelling through the Sun,
so the energy spread of a neutrino beam while traversing solar matter is left
unaltered. If neutrinos are endowed with magnetic moment, an idea originally
proposed by Cisneros \cite {Cisneros} and later revived by Voloshin, Vysotsky
and Okun \cite {VVO}, they can be converted via spin flip while travelling
through the solar magnetic field into weakly sterile particles. Due to their 
neutrality the direction of propagation is undeflected.

One of the original motivations for this hypothesis was the apparent 
anticorrelation of neutrino flux in the Homestake experiment with sunspot
activity \cite{RCD}. The data seemed to suggest that the more and the larger
the sunspots, the fewer neutrinos were detected, indicating a more efficient
spin flip conversion at times of more intense solar activity. Although this
is an attractive hypothesis, several statistical analyses \cite{BFP} - 
\cite{SSRD} fail to confirm this possibility and the data from all experiments
\cite{Hom} - \cite{SuperK} appear to be consistent with a constant rate 
\cite{W}.

It is now widely acknowledged that there exists a solar neutrino deficit
which, in very general terms, may be stated as too few neutrinos being 
detected \cite {Hom} - \cite{SuperK} as compared to theoretical predictions
\cite {TCL} - \cite{BP95}. The possibility of a solution to this discrepancy 
within the minimal standard electroweak model appears highly unlikely. Hopes
for solutions based on wrong interpretations of the experiments \cite {Hom} -
 \cite{SuperK} or different input solar physics have successively been 
abandoned \cite {Bahcallnu96}. More and more the right solution seems to lie 
on non-standard neutrino properties such as oscillations (both MSW \cite{MSW}
or vacuum \cite {HPS}), the resonant spin flavour precession mechanism (RSFP)
\cite {LMA} or flavour changing neutral current interactions \cite {ref20}. 
The RSFP mechanism combines the original idea of Cisneros \cite {Cisneros}
with the interesting possibility of the conversion mechanism being a resonant
one, like matter oscillations \cite {MSW} and is totally consistent, as will
be seen, with the absence of the above mentioned anticorrelation.

The present paper deals with RSFP type solutions to the solar neutrino 
problem. This mechanism requires a smaller magnetic field than the simple 
spin flip and one of its features is the fact that neutrinos of different
energies have their resonances in different locations within the Sun, an 
essential point in the present analysis.

In the past few years of solar neutrino physics, it has become increasingly 
clear that the deficit is energy dependent \cite{KR} - \cite{Berezinsky}, in 
the sense that weakly active neutrinos of different energies are suppressed 
differently. Since the resonance range is the zone where most if not all the 
conversion occurs and a stronger magnetic field leads to a greater 
suppression, an energy range where neutrinos are largely suppressed is a
clear indication of a large magnetic field in a certain range of the Sun's
interior. Conversely, a small suppression in a given energy range indicates a
small magnetic field in the region where the corresponding resonance is 
located. This principle is the basic idea of the present work and in this way
a 'tomography' of the solar magnetic field may be obtained. Hence the
neutrino energy suppression spectrum is expected to be directly connected to
the profile of the magnetic field along the Sun's radius, provided the 
neutrinos have a sizeable magnetic moment. It should be stressed that a 
sensible solution for the field profile is reasonably constrained, as it must
account for a strong suppression for the intermediate energy neutrinos, a 
moderate one for the energetic $^8B$ ones and no suppression for the $pp$ 
ones. This result together with the requirement that all neutrinos should have 
a resonance inside the Sun implies, as an essential feature, that the average 
magnetic field should undergo a discontinuity or at least a very fast 
variation in space somewhere in the interior of the Sun.

Recent progress in helioseismology seems to confirm the fact that the fluid 
character of the convective zone is interrupted near its bottom at 
approximately $ 0.71 R_{S}$. Below this depth a rigid body like 
structure appears. This is thus the likeliest region for the fast variation 
to arise. From the analysis of this paper one can predict that, starting from
the solar surface where the field should not exceed a few kG in the sunspots,
one expects an increase along the convective zone towards a maximum near its
bottom. This increase appears to become more intense in the deeper layers of 
the convective zone. Proceeding inwards in the radial direction, a sudden drop 
by an order of magnitude or more over the following $7 - 10 \% R_{S}$ is 
expected. Below this range the field is either negligible or may rise very 
moderately. This expectation is consistent with a neutrino magnetic moment in the 
range from a few times $10^{-13} \mu_{B}$ up to its laboratory bound of 
$10.8\times10^{-10}\mu_{B}$ \cite {LAMPF}.

The paper is divided as follows: in section 2 a brief review of the 
experimental situation \cite{Hom} - \cite{SuperK} is presented and its 
implications for the survival probabilities of low, intermediate and high
energy neutrinos are derived using the predictions from several standard
solar models \cite{TCL} - \cite{BP95}. In section 3 the neutrino 
propagation equation through solar matter is solved analytically and 
generalized to include the neutrino energy \cite{BU88} and production zone
distributions \cite{BP95} and non-adiabatic effects in the form of the Landau
Zener approximation \cite {LZ}. The expression of the survival probability is
next evaluated. In section 4 the necessary general profile which is 
required in order to satisfy experimental data and the results from solar 
models is established. It is argued that the high suppression found for the 
intermediate energy neutrinos and the absence of suppression for the 
low energy ones along with the fact that their energies are relatively close
is an indication of the sudden field variation referred to above. An example
of a toy magnetic field is obtained from approximate general conditions. In
section 5 an outlook is presented. Majorana neutrinos with transition 
magnetic moments ($\nu_{e}\rightarrow\bar{\nu}_{\mu}$) will be used 
throughout, the results for Dirac neutrinos being practically identical 
\cite {PhysRep}.

\vspace{10 mm}

{\bf \large  2. Energy Dependent Solar Neutrino Suppression}

\vspace{6 mm}

The data from the four solar neutrino experiments \cite{Hom} -\cite{SuperK}
are quoted in table I and the corresponding predictions from six solar models
\cite{TCL} - \cite{BP95} including the contribution from each neutrino 
component are given in table II. All models include heavy element 
diffusion except for TCL \cite{TCL} and TCCCD \cite{TCCCD}. 
It is now generally acknowledged that a 'standard'
solar model (SSM) should include diffusion, owing to the fact that such models 
give a remarkably good agreement with data from helioseismology \cite {IDFR}.

The absence of the intermediate energy neutrinos, consisting principally of 
the $^7Be$ line at $E=0.86MeV$ and the CNO continuum, has been realized 
several years ago \cite{Bahcall94} from the comparison of the Homestake and
Kamiokande data. It is considered by some as the 'true' solar neutrino problem
in the sense that it is independent from normalization to any solar model, 
either standard or non-standard \cite{DS96}. It appears as a natural 
consequence of the luminosity constraint \cite{BP95} , \cite{CIFLR} 
($L_{\odot}=1.367\times10^{-1}Wcm^{-2}$)

\be
L_{\odot}=\sum_{k}(\frac{Q}{2}-\!<\!E_{\nu}\!>_k\!)f_{k}~~~(k=pp,pep,
^7\!\!{Be},CNO,^8\!\!{B})
\ee
with $Q=26.73MeV$ (total energy released in each neutrino pair production) and
the equations \cite {CIFLR}
\bea
S_{Ga}&=&\sum_{i} \sigma_{Ga,i}f_{i}~~~(i=pp,pep,^7\!\!{Be},CNO,^8\!\!{B})\\
S_{Cl}&=&\sum_{j}\sigma_{Cl,j}f_{j}~~~(i=^7\!\!{Be},CNO,^8\!\!{B})\\
f_{pep}&=&0.0023f_{pp}
\eea
where $S_{Ga},S_{Cl}$ denote the Gallium and Chlorine data.

One has in fact from this system of four equations, upon elimination of the 
$pp$ flux $f_{pp}$, using the average value $\bar S_{Ga}=73.8\pm 7.7SNU$
and the nuclear cross sections \cite {CIFLR} 
$\sigma_{Ga,i}, \sigma_{Cl,j}$, the following intermediate energy neutrino 
flux (in units $cm^{-2}s^{-1})$
\bea
f_{Be}&=&1.04\times10^{4}f_{B}-2.88\times10^{10}\\
f_{CNO}&=&-8.46\times10^{3}f_{B}+2.22\times10^{10}.
\eea
Inserting $f_{B}$ from SuperKamiokande, \cite{SuperK} the total flux from 
these neutrinos is negative:
\bea
f_{Be}&=&-3.42\times10^{10}cm^{-2}s^{-1}\\
f_{CNO}&=&1.56\times10^{10}cm^{-2}s^{-1}.
\eea
Better fits were done by the authors of \cite{CIFLR},\cite{HR} who obtained
\bea
f_{Be+CNO}&\leq&0.7\times10^{9}cm^{-2}s^{-1}~~~~~(3\sigma)\\
f_{Be+CNO}&=&(-2.5\pm1.1)\times10^{9}cm^{-2}s^{-1}.
\eea
These authors used the former Kamiokande flux data which were higher than
Super-Kamiokande. From equations (5) and (6) it is seen that the total flux
$f_{Be+CNO}$ decreases with decreasing $f_{B}$, so that the results (9),(10)
should be further aggravated in the non-physical direction. Hence the 
probability that neutrinos are standard is no greater than 1\%, while, if the
luminosity constraint is dropped, it may increase to 4\% \cite {HR}. So 
intermediate energy neutrinos appear in practice to be completely suppressed.

As far as high energy ($^8B$) neutrinos are concerned and denoting by $R_{Cl}$
the ratio data/SSM prediction one may write
\be
R_{Cl}=R^{I}_{Cl}P_{I}+R^{H}_{Cl}P_{H}.
\ee
Here $R^{I(H)}_{Cl}$ is the fraction of intermediate (high) energy neutrinos 
in the Chlorine experiment as theoretically predicted and $P_{I(H)}$ is the
fraction of intermediate (high) energy $\nu_{e_L}$ produced in the Sun that
are detected on Earth. Using $P_{I}=0$ (99\%CL) and the models listed in 
table I one gets for $P_{H}$ the range
\be
0.35<P_H<0.63
\ee
with the smaller value corresponding to BP95 \cite{BP95}and the larger to 
TCCCD \cite{TCCCD}.

The same argument may now be applied to the low energy ($pp$) neutrinos. With
obvious notation, one has for the Gallium result
\be
R_{Ga}=R^{L}_{Ga}P_{L}+R^{I}_{Ga}P_{I}+R^{H}_{Ga}P_{H}
\ee
Again, taking $P_{I}=0$, one obtains for the two extreme cases
\bea
P_{L}&=&0.981~~~~~(\rm{BP}95) \cite{BP95}\\ 
P_{L}&=&0.966~~~~~(\rm{TCCCD}) \cite{TCCCD}
\eea
which means that the low energy ($pp$) neutrinos are practically unsuppressed.
Thus any solution to the solar neutrino problem based on non-standard 
neutrino properties must be consistent with the following general 
conclusions:\\
1. No suppression for low energy ($pp$) neutrinos.\\
2. Strong suppression for intermediate energy ($^7Be$, CNO) neutrinos.\\
3. Moderate suppression for high energy ($^8B$) neutrinos.\\
The requirements $P_{L}>0.96$, $P_{I}<0.04$, $0.35<P_{H}<0.63$ will be used 
throughout the paper.

\vspace{10mm}

{\bf \large 3. Survival Amplitudes and Probabilities}

\vspace{6mm}

The starting point of this section is the Schr\"{o}dinger-like evolution 
equation for Majorana neutrinos in the Sun \cite {PhysRep}
\be
-i\frac{d}{dr}\left(\begin{array}{c}\nu_{e_L}\\
\bar{\nu}_{\mu_R}
\end{array}\right)=H\left(\begin{array}{c}\nu_{e_L}\\
\bar{\nu}_{\mu_R}
\end{array}\right)
\ee
with the Hamiltonian H given by
\be
H=\left(\begin{array}{cc}
(\frac{G}{\sqrt{2}})\frac{11}{6}N_e&{\mu}B\\
{\mu}B&\frac{\Delta^2m_{21}}{2E}+(\frac{G}{\sqrt{2}})\frac{1}{6}N_e
\end{array}\right).
\ee
Here $\Delta^2m_{21}$ is the mass square difference between mass eigenstates,
$N_e$ is the electron density taken as exponentially decreasing with radius in 
the region of interest \cite{BU88}, $\!GN_e\!=\!2.11\!\times\!10^{-11}\!
exp(-r/0.09R_S)eV\!$
, B is the solar magnetic field and the remaining notation is obvious. A 
vanishing vacuum mixing angle is assumed throughout and the mixing angle from
matter and magnetic moment/field is \cite{PhysRep}
\be
{s^2}_\theta=\frac{\mu^2B^2}
{\mu^2B^2+\frac{1}{4}[\frac{5G}{3\sqrt{2}}N_e-\frac{\Delta^2m_{21}}{2E}+
((\frac{5G}{3\sqrt{2}}N_e-\frac{\Delta^2m_{21}}{2E})^2+4\mu^2B^2)^{1/2}]^2}.
\ee
The neutrino propagation through solar matter is assumed to be adiabatic 
except for a short range \cite{PhysRep} in the neighbourhood of a critical 
point - the resonance - where the mixing angle (18) is maximal so that the
density verifies
\be
\frac{5G}{3\sqrt{2}}N_e=\frac{\Delta^2m_{21}}{2E}.
\ee
Across the resonance the neutrinos may shift from one eigenvalue to another,
this effect being parametrized by a jump amplitude. So a neutrino produced as
a weak eigenstate $\nu_{e_L}$ at $r_i$ which is a superposition of two mass 
eigenstates $\nu_1,\nu_2$
\be
\left(\begin{array}{c}\nu_1(r_i)\\
\nu_2(r_i)
\end{array}\right)=\left(\begin{array}{cc}
c_{\theta_{i}}&s_{\theta{i}}\\
-s_{\theta_{i}}&c_{\theta{i}}
\end{array}\right)
\left(\begin{array}{c}1\\
0
\end{array}\right)
\ee
will become at location $r_1$, after having passed through a resonance 
\cite {Parke}, the following combination
\be
\left(\begin{array}{c}\nu_1(r_1)\\
\nu_2(r_1)
\end{array}\right)=
\left(\begin{array}{cc}
\lambda_{1}exp~i\int_{r_i}^{r_1}H_{D_{11}}dr&\lambda_{2}exp~i\int_{r_i}^{r_1}
H_{D_{22}}dr\\
-\lambda^{*}_{2}exp~i\int_{r_i}^{r_1}H_{D_{11}}dr&\lambda^{*}_{1}exp~i
\int_{r_i}^{r_1}
H_{D_{22}}dr
\end{array}\right)
\left(\begin{array}{c}\nu_1(r_i)\\
\nu_2(r_i)
\end{array}\right).
\ee
Here $\lambda_1$ denotes the amplitude for adiabatic propagation, $\lambda_2$ 
is the jump amplitude across the resonance and $H_{D_{11}}$, $H_{D_{22}}$
are the 11 and 22 elements of the diagonalized Hamiltonian. Denoting by 
$\theta_1$ the diagonalizing angle at $r_1$, it is 
straightforward to derive the survival amplitude, that is, the amplitude that
a neutrino which is produced as $\nu_{e_L}$ remains as $\nu_{e_L}$
\bea
Amp(\nu_{e_L}(r_i)\rightarrow\nu_{e_L}(r_1))=c_{\theta_1}c_{\theta_i}\lambda_1
exp~i\!\!\int_{r_i}^{r_1}\!\!H_{D_{11}}dr-c_{\theta_1}s_{\theta_i}\lambda_2
exp~i\!\!\int_{r_i}^{r_1}\!\!H_{D_{22}}dr \nonumber \\
+s_{\theta_1}c_{\theta_i}\lambda_2^{*}exp~i\!\!
\int_{r_i}^{r_1}\!\!H_{D_{11}}dr+s_{\theta_1}s_{\theta_i}\lambda_1^{*}
exp~i\!\!\int_{r_i}^{r_1}\!\!H_{D_{22}}dr. \nonumber\\
\eea
For the conversion amplitude one just performs in (22) the replacements
$c_{\theta_1}\rightarrow{s_{\theta_1}}$, 
$-s_{\theta_1}\rightarrow{c_{\theta_1}}$.
In a simplified calculation where all neutrinos are considered to be emmited
at the same point and with the same energy, the survival probability is simply 
the square modulus of eq. (22) where all fast oscillating terms are set equal 
to zero. This is the well known Parke's formula \cite{Parke}
\be
P_{surv}=\frac{1}{2}+(\frac{1}{2}-P_C)c_{2\theta_i}c_{2\theta_1}
\ee
where the quantity $P_C$ is the jump probability, $P_C=|\lambda_2|^2$, which 
will be taken in the Landau Zener approximation \cite{LZ,PRD} \footnote{It
was previously shown that the Landau Zener approximation which assumes a 
linearly decreasing density in the vicinity of the critical point works rather
well in the Sun. See \cite{PRD}}.
\be
P_C=exp(-\pi\frac{2\mu^2B^2}{\frac{\Delta^2m_{21}}{2E}}0.09R_S).
\ee
The coefficient of this exponent is the adiabaticity parameter,
\be
\gamma_C=\frac{2\mu^2B^2}{\frac{\Delta^2m_{21}}{2E}}0.09R_S.
\ee

A more accurate calculation must include the production and energy range of 
the neutrinos. Denoting by $\chi(E)$ and $\phi(r_i)$ the production amplitudes 
per unit energy and unit length respectively, the survival amplitude at the 
edge of the Sun ($r_1=R_S$) becomes the sum of the following four terms
\newpage
\be
\nu_e=\int_{E_m}^{E_M}\!\chi(E)c_{\theta_1}\!\int_{a}^{b}
\!\lambda_1\phi(r_i)c_{\theta_i}exp~i\!\int_{r_i}^{R_S}\!
H_{D_{11}}drdr_{i}dE \nonumber\\
\ee
\be
-\!\int_{E_m}^{E_M}\!\chi(E)
c_{\theta_1}\!\int_{a}^{b}\!\lambda_2\phi(r_i)s_{\theta_i}exp~i
\!\int_{r_i}^{R_S}\!H_{D_{22}}drdr_{i}dE \nonumber\\
\ee
\be
+\!\int_{E_m}^{E_M}\!\chi(E)s_{\theta_1}\!\int_{a}^{b}\!
\lambda^{*}_{2}\phi(r_i)c_{\theta_i}exp~i\!\int_{r_i}^{R_S}\!
H_{D_{11}}drdr_{i}dE \nonumber\\
\ee
\be
+\!\int_{E_m}^{E_M}\!
\chi(E)s_{\theta_1}\!\int_{a}^{b}\!\lambda^{*}_{1}\phi(r_i)
s_{\theta_i}exp~i\!\int_{r_i}^{R_S}\!H_{D_{22}}drdr_{i}dE. 
\ee
while the conversion amplitude
\be
\bar{\nu}_{\mu}=\int_{E_m}^{E_M}\!\chi(E)s_{\theta_1}\!\int_{a}^{b}
\!\lambda_1\phi(r_i)c_{\theta_i}exp~i\!\int_{r_i}^{R_S}\!
H_{D_{11}}drdr_{i}dE \nonumber\\
\ee
\be
-\!\int_{E_m}^{E_M}\!\chi(E)
s_{\theta_1}\!\int_{a}^{b}\!\lambda_2\phi(r_i)s_{\theta_i}exp~i
\!\int_{r_i}^{R_S}\!H_{D_{22}}drdr_{i}dE \nonumber\\
\ee
\be
-\!\int_{E_m}^{E_M}\!\chi(E)c_{\theta_1}\!\int_{a}^{b}\!
\lambda^{*}_{2}\phi(r_i)c_{\theta_i}exp~i\!\int_{r_i}^{R_S}\!
H_{D_{11}}drdr_{i}dE \nonumber\\
\ee
\be
-\!\int_{E_m}^{E_M}\!
\chi(E)c_{\theta_1}\!\int_{a}^{b}\!\lambda^{*}_{1}\phi(r_i)
s_{\theta_i}exp~i\!\int_{r_i}^{R_S}\!H_{D_{22}}drdr_{i}dE. 
\ee

In these expressions the first two integrals in each term extend over the 
energy and production range for each neutrino component ($pp$,$^7Be$,$^8B$,
...) as given in refs. \cite{BU88}, \cite{BP95} respectively (model with $He$ 
and metal diffusion).

Turning now to the probabilities, it should first be observed that nuclei 
produce neutrinos in a disordered manner both in space-time and in energy.
Each neutrino therefore acquires a random phase endowing the terms (26) - (29)
and (30) - (33) which therefore add incoherently, as opposed to (22). The 
random neutrino emission thus makes the amplitudes completely incoherent, so 
one adds probabilities. The generalization of (23) including energy and 
production range distribution is thus
\be
P_{surv}=\!\int_{E_m}^{E_M}\!\!g(E)\!\int_{a}^{b}\!\!f(r_i)~\left[\frac{1}{2}+
(\frac{1}{2}-P_C)\right]~c_{2_{\theta_1}}c_{2_{\theta_i}}dr_{i}dE.
\ee
For the conversion probability one just performs in (34) the replacement $(+) 
\rightarrow (-)$. The functions $g(E), f(r_i)$, representing respectively the 
probability density of neutrino production per unit energy and per unit 
length, will be taken from the numerical tables in refs. \cite{BU88}, 
\cite{BP95}. An example of a survival probability showing the characteristic
'bath tub' shape is shown in fig. 1.

Equations (34), (24), (25) constitute the basis of the following analysis 
in section 4.

\newpage

{\bf \large 4. Magnetic Field Profiles}

\vspace{4mm}

{\bf 4.1 General Features}

\vspace{4mm}

Apart from the knowledge that neutrinos may provide us if they have a sizeable
magnetic moment, very little is known about the solar magnetic field. From the
observation of the interstellar field, a lower bound may be fixed at $10{\mu}G$,
whereas an upper bound of $(2-5)\times10^3G$ near the surface may be 
considered as associated to the field in sunspots. There is however no 
indication as to the extension of these structures and its associated field in
depth significantly below the surface layers.

Several 'plausible' profiles have been proposed and investigated in the 
litterature \cite{ALP} - \cite{BHL} but these analyses which are now several 
years old
did not take into account an energy dependent solar neutrino suppression. It
should first be emphasized that the resonance condition (19) fixes the 
location of the critical density for each neutrino according to its own 
energy \cite {PhysRep}:
\be
x_{res}=0.09log\frac{\frac{5}{3\sqrt{2}}~2.11\!\!\times\!\!10^{-11}eV}
{\frac{\Delta^2m_{21}}{2E}}
\ee
where $x$ is the fraction of the solar radius $x=r/R_S$.

The range around the critical density where the Hamiltonian eigenvalue 
splitting,
\be
{\alpha}_1-{\alpha}_2=\sqrt{\left(\frac{5G}{3\sqrt{2}}N_e-
\frac{\Delta^2m_{21}}{2E}\right)^2+4{\mu}^2B^2}~~,
\ee
is dominated by the term $4{\mu}^2B^2$ is where most if not all the conversion
(i. e. neutrino suppression) occurs.

An efficient conversion means that the jump probability (24) across the 
resonance from one Hamiltonian eigenvalue to the other is small, so each
neutrino remains in its own eigenvalue. The process is therefore adiabatic
and $\gamma_C$ (25) is large. Most of the neutrinos will gradually change 
their flavour and the space rate of this change is maximal at the vicinity of 
the critical point. Conversely, a sizeable jump probability means that a 
significant fraction of neutrinos will change from one Hamiltonian eigenvalue
to the other, thereby staying in the same weak eigenstate and the process is
non-adiabatic. In all resonant processes the resonance is of course located
between the production zone and the surface of the Sun, so the mixing angle 
(18) at the production point is close to zero and at the edge of the Sun is
close to ${\pi}/2$ \cite {PhysRep}. In the rough approximation (23) with no
energy range nor production zone distribution, this means 
$c_{2\theta_i}\simeq1$ and $c_{2\theta_1}\simeq-1$ so that
\be
P_{surv}{\simeq}P_C.
\ee

A large jump probability opposes conversion, thus favouring survival. In such 
case most neutrinos, while travelling across the resonance, will leave to the 
other Hamiltonian eigenvalue which corresponds, past the resonance, mainly to 
the initial weak eigenstate. So neutrino survival is related to strongly 
adiabatic processes and conversion to the lack of adiabaticity. Recalling the 
analysis of section 2, these arguments show that $pp$ neutrinos, for which
$P_L{\geq}0.96$, must correspond to highly non-adiabatic resonances,
\be
{\gamma}_C{\leq}0.01
\ee
while intermediate energy neutrinos which are highly suppressed 
($P_I{\leq}0.04)$ must correspond to 
\be
{\gamma}_C{\geq}1.
\ee
Requiring condition (38) to hold for all $pp$ neutrinos whose energies lie in
the interval
\be
0.236MeV{\leq}E_{pp}{\leq}0.420MeV
\ee
and taking intermediate energy neutrinos satisfying (39) to consist totally
of the $^7Be$ line at
\be
E_{^7\!Be}=0.86MeV
\ee
one has respectively from (38), (39) using (25)
\bea
0.09R_S2{\mu^2}B^2(x_{^7\!Be})~{\geq}~\frac{\Delta^2m_{21}}{2E_{^7\!Be}}\\
0.09R_S2{\mu^2}B^2(x_{pp})~{\leq}~0.01\frac{\Delta^2m_{21}}{2E_{pp}}.
\eea
The solar magnetic field should satisfy these two conditions at the 
corresponding critical densities. Using the values for $E_{^7\!Be}=0.86MeV$ 
and $E_{pp_{max}}=0.42MeV$, it is seen from (35) that the spacing between 
these critical densities is
\be
{\Delta}x_{^7\!Be,pp}=(0.065 - 0.116)
\ee
implying a sharp rise of the magnetic field by at least a factor of (6 - 7)
over a relatively short radial distance. Therefore it is seen from (42), (43)
that the field is both bound from below and from above at points whose mutual
distance is ${\Delta}x_{^7\!Be,pp}$:
\bea
B(x_{^7\!Be})~{\geq}~\frac{5.22\times10^{-3}}{f}\sqrt{\Delta^2m_{21}} \\
B(x_{pp})~{\leq}~\frac{0.483}{f}\sqrt{\frac{\Delta^2m_{21}}{E_{pp}}}  .
\eea
Here f is the neutrino magnetic moment in Bohr magnetons, 
${\mu}_{\nu}=f{\mu}_B$, energies are in eV and B in Gauss. No information is 
provided from this analysis as to the location of this particular topology 
of the solar magnetic field. Helioseismology indicates that such a behaviour 
can only appear around the bottom of the convective zone at approximately
$0.71R_S$. This constrains, as seen from (35), the parameter $\Delta^2m_{21}$ 
to lie within less than an order of magnitude from $10^{-8}eV^2$.

As far as ${^8}$B are concerned, their spread in energy is much larger than 
the previous ones \cite{BU88}
\be
0.814MeV<E_{^8\!B}<15MeV
\ee
so some of their resonances overlap with the one from $^7Be$. Given the 
survival probability expected for these neutrinos as derived in section 2
\be
0.35<P_H<0.63 ,
\ee
one may naively expect from (37), throughout most of their critical density
range, ${\gamma}_C$ to be situated in the interval
\be
{\gamma}_C{\simeq}(0.15 - 0.30).
\ee
However, due to the large energy spread here, no definite general criterion
can be established for the field distribution along the critical density 
range. For a short interval at the beginning of this range owing to the 
overlap with the $^7Be$ resonance, the field should of course be constrained
as discussed, but for the remainder, the actual profile is essentially 
unconstrained, except that the field should be substantially smaller on
average. Only the total probability (12) can be used as an indicator, together
with the requirement that the surface field should not exceed a few kG. It 
may remain quite large over a considerable extent of the critical density
range, so that an almost total conversion will occur for a substantial 
fraction of $^8B$ neutrinos with $\gamma_C$ far exceeding in this zone the 
interval (49). Consequently, in the remaining part, the field should be much
smaller with highly non-adiabatic transitions and $\gamma_C$ well below (49).
Such is the case with one of the distributions to be examined below. So the
requirement that $\gamma_C$ lies within the interval (49) for the resonances
of $^8B$ neutrinos is an indicative one for distributions varying smoothly in
the range $x>x_{^7\!Be}$.

\vspace{6mm}

{\bf 4.2 Numerical Examples}

\vspace{4mm}

The general characteristics of suitable magnetic fields expressed in 
inequalities (45), (46) will now be confronted with specific field 
distributions. Equation (34) for the survival probability will be used 
together with the requirements $P_I<0.04$, $P_L>0.96$ and $0.35<P_H<0.63$ as
discussed in section 2.

The starting point is a field spread only over the convective zone proposed
years ago \cite{BHL}. It changes from zero at $0.7R_S$ to $10^5$ G at 
$0.75R_S$, then remains constant until $0.8R_S$ and falls linearly to 
$2\times10^3$ G at the surface of the Sun. For simplicity, a linear increase
from $0.7R_S$ will also be assumed:
\be
\begin{array}{cclc}
B(x) & = & 0                         & 0{\leq}x{\leq}0.7 \\
B(x) & = & 2\times10^6(x-0.7)        & 0.7{\leq}x{\leq}0.75 \\
B(x) & = & 10^5                      & 0.75{\leq}x{\leq}0.8 \\
B(x) & = & 10^5-4.9\times10^5(x-0.8) & 0.8{\leq}x{\leq}1  . 
\end{array}
\ee

It is straightforward to realize that this field satisfies conditions (45),
(46). Requiring in fact (45), (46) to hold simultaneously for the $^7Be$ and 
all $pp$ resonances, i. e. up to $E_{pp_{max}}=0.42 MeV$, the intensity of the 
field in which $pp$ resonances are located is at most 1/7 of the field at the 
$^7Be$ resonance. Since the distance between the critical densities 
corresponding to $E_{pp_{max}}$ and $E_{^7Be}$ is $0.065R_S$, (see eq. (35)), 
this leads, using (50), to
\be
x_{^7\!Be}-0.065{\leq}0.707
\ee
(i. e. all $pp$ resonances should be located deeper than $0.707R_S$) and
\be
x_{^7\!Be}>0.700 ,
\ee
this last condition to ensure a non-vanishing magnetic field at the $^7Be$
resonance. Inequalities (51), (52) imply using (28)
\be
8.0\times10^{-9}eV^2{\leq}\Delta^2m_{21}{\leq}1.8\times10^{-8}eV^2 .
\ee
This double inequality thus reflects the compatibility between the $pp$ and 
$^7Be$ solutions in (50) according to conditions (45), (46). No constraint 
from the $^8B$ flux probability has yet been imposed. Using the lower bound
of (53) in (45) with the corresponding critical density field 
$B(x_{^7Be})=10^5G$ to determine within this criterion the lowest neutrino
magnetic moment compatible with $pp$ and $^7Be$ solutions, one obtains
\be
{\mu_{\nu}}\geq4.7\times10^{-12}{\mu_{B}} .
\ee
Using this value in eq. (34) for the survival probability with $P_C$ as in
(24) one finds, however, a minor overlap in $\Delta^2m_{21}$ as regards $pp$
and $^7Be$ neutrinos:
\be
\begin{array}{lll}
^8B  & (4.0 - 5.95\times10^{-9})eV^2              & (9.0\times10^{-8} - 1.25
\times10^{-7})eV^2 \\
^7Be & (1.2\times10^{-9} - 1.13\times10^{-8})eV^2 &   
                   \\
pp   & <4.3\times10^{-11}eV^2                     & >6.85\times10^{-9}eV^2 .
                   \\
\end{array}
\ee
This overlap $[\Delta^2m_{21}=(0.69 - 1.1)\times10^{-8}eV^2]$ is due to the 
fact that (45), (46) were imposed in this example for all $pp$ neutrinos which
need not be the case. A small fraction of $pp$ neutrinos may have their 
resonances in a very short range with a field slightly in excess of 1/7 of the
field at the $^7Be$ resonance, provided a substantial fraction of $pp$ 
resonances is located in a range where the field is much smaller. In terms of
$x_{pp}$ this overlap range where (45), (46) are not satisfied is
\be
0.707<x_{pp}<0.721
\ee
and for smaller $x_{pp}$ the field (50) is of course below 1/7 of $B(^7Be)$.
On the other hand the solution for $^8B$ neutrinos in (55) is clearly 
inconsistent with the other two. It can be turned however consistent by 
relaxing (45), (46) over a short range, hence relaxing (54). In this way the
set of solutions allowed by (50) may be investigated. The result is
\be
3.1\times10^{-12}{\mu_B}{\leq}{\mu_{\nu}}{\leq}3.8\times10^{-12}{\mu_B}~,~ 
6.33\times10^{-9}eV^2{\leq}\Delta^2m_{21}{\leq}6.56\times10^{-9}eV^2 
\ee
respectively.
In order to have an idea of the amount of maximum deviation from (45), (46)
involved, take the 'worst' solution, namely
\be
\mu_{\nu}=3.1\times10^{-12}\mu_B~~~,~~~\Delta^2m_{21}=6.33\times10^{-9}eV^2 .
\ee
Condition (45) then gives $B(x_{^7Be}){\geq}133970G$ (actually 
$B(x_{^7Be})=10^5G$) while (46) is satisfied for all neutrinos with energies
$E{\leq}0.342MeV$. The field distribution (50), shown in fig.2, is therefore 
a borderline case in terms of providing a solution to the solar neutrino 
problem. 

What one learns from this example is that imposing inequalities (45), (46)
over the whole $pp$ resonance range provides an approximate sufficient 
condition
for a suitable field distribution. Minor deviations may be allowed for some
$x_{pp}$, since these inequalities were derived from the simplified formula 
(37) and a fixed neutrino energy. In any case, though, the main characteristic
of a field increasing by a factor $>6$ over a fraction shorter than 9\% of the
solar radius seems essential, in order for the field to be compatible with
solar neutrino data: the smearing effect due to the energy distribution is
not as large with $pp$ as with $^8B$ neutrinos, so this criterion should be
retained.

Let us reverse the attitude and search for a field distribution satisfying 
(45), (46) behaving linearly with $x$ for simplicity. The $^7Be$ resonance is 
assumed to be located at the bottom of the convective zone where the field is
taken to be maximal ($x_{Be}=0.71$, $\Delta^2m_{21}=1.6\times10^{-8}eV^2$).
It then falls to $2\times10^3G$ at the surface of the Sun. Taking 
$f=10^{-12}$, one has $B(x_{^7\!Be}){\geq}6.6\times10^5G$ and (46) becomes
\be
B(x_{pp})\leq6.11\times10^7\frac{1}{\sqrt{E_{pp}}}G
\ee
with $E_{pp}$ given in eV. Given the fact that $pp$ resonances start at 
$0.065$ below $x_{^7\!Be}$, one has essentially two kinds of options:

1. First option. One imposes inequality (46) to hold only for part of $pp$
resonances $(x_{pp}{\leq}x_{^7\!Be}-{\Delta}x)$ thus allowing for its 
violation in the interval 
$x_{^7\!Be}-{\Delta}x{\leq}x_{pp}{\leq}x_{^7\!Be}-0.065$. Consequently $B(x)$
has to further decrease inwards from its value at 
$x_{pp}=x_{^7\!Be}-{\Delta}x$ in order to compensate for excessive neutrino 
conversion in this interval. (Recall that $pp$ neutrinos are practically 
unconverted). So, taking for instance ${\Delta}x=0.09$, the
corresponding energy $E_{pp}$ in (46) for 
${\Delta}^2m_{21}=1.6\times10^{-8}eV^2$ is $3.17\times10^{5}eV$. This gives
$B(x_{pp}){\leq}1.08\times10^5G$ from (59) ($B(x_{pp}){\leq}9.4\times10^4G$
if one required all $pp$ neutrinos to satisfy (46), i. e. ${\Delta}x=0.065$).
Hence, choosing $B(x_{pp}){\leq}9.4\times10^4G$ to allow for an overlap, the
field will be
\be
\begin{array}{cclc}
B(x) & = & 0                                    & x{\leq}0.606 \\
B(x) & = & 6.73\times10^6(x-0.62)+9.4\times10^4 & 0.606{\leq}x{\leq}0.71 \\
B(x) & = & -2.407\times10^6(x-0.71)+7\times10^5 & 0.71{\leq}x{\leq}1 .\\
\end{array}
\ee
However the probability constraints on $P_{H}$, $P_{I}$, $P_{L}$ give 
respectively
\be
\begin{array}{lll}
^8B  & (4.8 - 6.8)\times10^{-9}eV^2               & 
(1.95 - 2.70)\times10^{-7}eV^2 \\
^7Be & (1.12\times10^{-9} - 2.1\times10^{-8})eV^2 &           \\
pp   & <8.5\times10^{-11}eV^2                     & 1.1\times10^{-8}eV^2 ,
\end{array}
\ee
with ${\mu_{\nu}}=10^{-12}{\mu_B}$, showing an inconsistency with $^8B$. This 
is solved by requiring a more moderate slope of the field along the outer 
layers of the convective zone. A decrease from  $2.2\times10^5G$ to
$2\times10^4G$ at $x=0.91$ is sufficient while keeping the same values at the
surface and $x_{^7Be}=0.71$. Thus the same suppression of $^8B$ neutrinos is
achieved at larger densities than those corresponding to the interval 
$(4.8 - 6.8)\times10^{-9}eV^2$ in (61). The field is then (see fig.3):
\be
\begin{array}{cclc}
B(x) & = & 0                                    & x{\leq}0.606 \\
B(x) & = & 6.73\times10^6(x-0.62)+9.4\times10^4 & 0.606{\leq}x{\leq}0.71 \\
B(x) & = & -3.4\times10^6(x-0.71)+7\times10^5   & 0.71{\leq}x{\leq}0.91 \\ 
B(x) & = & -2\times10^5(x-0.91)+2\times10^4     & 0.91{\leq}x{\leq}1 \\
\end{array}
\ee
with a solution ($f=10^{-12}$) \footnote{It is worth noting that the range of
possible neutrino magnetic moments extends in this case down to 
${\mu_{\nu}}=6.5\times10^{-13}{\mu_B}$.}
\be
\begin{array}{lll}
^8B  & (1.37 - 1.90)\times10^{-8}eV^2            & 
(1.95 -2.70)\times10^{-7}eV^2                                              \\
^7Be & (3.2\times10^{-9} - 2.1\times10^{-8})eV^2 &                         \\
pp   & <1.34\times^{-10}eV^2                     & >1.65\times10^{-8}eV^2 .\\
\end{array}
\ee
Hence ${\Delta}^2m_{21}=(1.65 - 1.90)\times10^{-8}eV^2$.

2. Second option. Inequality (46) is imposed as from $x_{^7Be}-0.065$ inwards.
B(x) may then ramain constant below this value since it will be sufficiently 
small throughout all $x_{pp}$ to ensure enough neutrino survival. Again a 
large slope at large $x$ (upper convective zone) must be avoided for 
consistency with $^8B$ neutrinos. So $B(x_{pp})=9.4\times10^4G$ for all 
$x_{pp}$ and, as before, $B(x_{^7Be})=7\times10^5G$, (see (fig.4)):
\be
\begin{array}{cclc}
B(x) & = & 9.4\times10^4                         &            x{\leq}0.645  \\
B(x) & = & 9.32\times10^6(x-0.645)+9.4\times10^4 & 0.645{\leq}x{\leq}0.71 \\
B(x) & = & -3.4\times10^6(x-0.71)+7\times10^5    & 0.71{\leq}x{\leq}0.91   \\
B(x) & = & -2\times10^5(x-0.91)+2\times10^4      & 0.91{\leq}x{\leq}1      \\
\end{array}
\ee
with a solution ($f=10^{-12}$) 
\be
\begin{array}{lll}
^8B  & (1.37 - 1.91)\times10^{-8}eV^2             & 
(1.73 - 2.37)\times10^{-8}eV^2  \\
^7Be & (3.2\times10^{-9} - 1.95\times10^{-8})eV^2 &      \\
pp   &  <1.34\times10^{-10}eV^2                    & >1.5\times10^{-8}eV^2.
\end{array}
\ee

The range of solutions is slightly enlarged here relative to option 1: 
$\Delta^2m_{21}=(1.50 - 1.91)\times10^{-8}eV^2$.

Such a large field at the bottom of the convective zone ($7\times10^5G$) is a
consequence of requiring a comparatively small value of the neutrino magnetic 
moment (${\mu_{\nu}}=10^{-12}\mu_B$) and, especially, the $^7Be$ resonance to
to lie at $0.71R_S$ instead of lower densities as in the previous example 
(50). For lower densities, adiabaticity is in fact satisfied with lower 
magnetic fields (see eq. (25)).

At this stage it should have become obvious that a monotonically decreasing
field from the core or the inner radiative zone to the solar surface such as
the ones proposed in refs. \cite{ALP,ALP1} must be ruled out in the light of 
solar neutrino data. Also linearly varying profiles, although not to be 
excluded, can only provide borderline solutions. 

To conclude this section, four more field profiles rising sharply at the 
bottom of the convective zone will be briefly reviewed. The first two are 
variants of one of the fields proposed in ref. \cite{ALP2} namely,
\be
B(x)=\frac{3.048\times10^4}{ch[20(x-0.71)]}G~~~,~~~0.71{\leq}x{\leq}1
\ee
with $B=2.8\times10^3G$ and $B=0$ for $x<0.71$ (see fig.5). For the first of 
these one obtains the following two limiting cases
\be
\begin{array}{lll}
^8B  & (1.60 - 2.50)\times10^{-8}eV^2            & 
(1.10 - 1.54)\times10^{-7}eV^2 \\
^7Be & (5.2\times10^{-9} - 1.6\times10^{-8})eV^2 &          \\
pp   & <7.4\times10^{-11}eV^2                    & >1.43\times10^{-8}eV^2 \\
\end{array}
\ee
with ${\mu}_{\nu}=3.4\times10^{-11}{\mu}_B$ and
\be
\begin{array}{lll}
^8B  & (1.50 - 2.40)\times10^{-8}eV^2            & 
(1.10 - 1.55)\times10^{-7}eV^2 \\
^7Be & (4.9\times10^{-9} - 1.6\times10^{-8})eV^2 &          \\
pp   & <6.9\times10^{-11}eV^2                    & >1.60\times10^{-8}eV^2
\end{array}
\ee
with ${\mu}_{\nu}=3.65\times10^{-11}{\mu}_B$. This leads to the range of 
solutions: 
\be
3.4\times10^{-11}{\mu_B}{\leq}{\mu_{\nu}}{\leq}3.65\times10^{-11}{\mu_B}~,~
1.55\times10^{-8}eV^2{\leq}{\Delta}^2m_{21}{\leq}1.60\times10^{-8}eV^2
\ee
with (46) satisfied for $x_{pp}{\leq}x_{^7\!Be}-0.082$. As opposed to the 
previous examples, the field is constant below its maximum, so the fraction of
$pp$ resonances that do not satisfy (46) has to be smaller. For the second of 
distributions (66), with the field limited to the convection zone, the range
of solutions necessarily extends. In fact the two limiting cases are now
\be
\begin{array}{lll}
^8B  & (1.60 - 2.50)\times10^{-8}eV^2             &
(1.0 - 1.42)\times10^{-7}eV^2 \\
^7Be & (5.1\times10^{-9} - 1.60\times10^{-8})eV^2 &   \\
pp   & <6.4\times10^{-11}eV^2                      & >7.3\times10^{-9}eV^2 
\end{array}
\ee
with ${\mu_{\nu}}=3.4\times10^{-11}{\mu_B}$ and
\be
\begin{array}{lll}
^8B  & (3.3 - 7.3)\times10^{-9}eV^2                &
(1.10 - 1.44)\times10^{-7}eV^2  \\
^7Be & (2.0\times10^{-9} - 1.6\times10^{-8})eV^2   &  \\
pp   &  no~solution                                & >7.3\times10^{-9}eV^2
\end{array}
\ee
with ${\mu_{\nu}}=1.3\times10^{-10}{\mu_B}$. The range of 
solutions is now
\be
3.4\times10^{-11}{\mu_B}{\leq}{\mu_{\nu}}{\leq}1.3\times10^{-10}{\mu_B}~,~
7.3\times10^{-9}eV^2{\leq}{\Delta}^2m_{21}{\leq}1.60\times10^{-8}eV^2.
\ee

The last two profiles to be examined bear no discontinuity. They are
\be
\begin{array}{cclc}
B(x) & = & 2.16\times10^3G                            & x{\leq}0.7105  \\
B(x) & = & 8.7\times10^4[1-\left({\frac{x-0.75}{0.04}}\right)^2]G & 
0.7105{\leq}x{\leq}0.7483 \\
\end{array}
\ee
with
\be
B(x)=10^5[1-3.4412(x-0.71)]G~~~~0.7483{\leq}x{\leq}1
\ee
and
\be
B(x)=\frac{8.684\times10^4}{ch[20(x-0.7483)]}G~~~~~~~~~~0.7483{\leq}x{\leq}1~
\ee
shown in figs.6 and 7 respectively.
The first of these has a very limited range of solutions of which some violate
up to 30\% the inequality (45). Also (46) is satisfied only for $E<0.381MeV$,
a similar situation to that encountered with the linear field distribution 
(50). Again, this is a borderline case and the limiting solutions are
\be
\begin{array}{lll}
^8B  & (6.2 - 9.1)\times10^{-9}eV^2 & (8.6\times10^{-8} - 
1.2\times10^{-7})eV^2 \\
^7Be & (3.5 - 6.5)\times10^{-9}eV^2 &       \\
pp   & <1.0\times10^{-10}eV^2       & >6.5\times10^{-9}eV^2
\end{array}
\ee
with ${\mu_{\nu}}=4.1\times10^{-12}{\mu_{B}}$ and
\be
\begin{array}{lll}
^8B  & (4.8 - 6.8)\times10^{-9}eV^2               & 
(9.5\times10^{-8} - 1.3\times10^{-7})eV^2  \\
^7Be & (1.3\times10^{-9} - 1.25\times10^{-8})eV^2 & \\
pp   & <9.2\times10^{-11}eV^2                     & >6.9\times10^{-9}eV^2
\end{array}
\ee
with ${\mu_{\nu}}=6.15\times10^{-12}{\mu_B}$. These lead to the following
range of solutions 
\be
4.1\times10^{-12}{\mu_B}{\leq}{\mu_{\nu}}{\leq}6.15\times10^{-12}{\mu}_B~,~
6.5\times10^{-9}eV^2{\leq}{\Delta^2m_{21}}{\leq}6.9\times10^{-9}eV^2~.
\ee
In contrast to (73), (74), the second combination (73), (75), provides a much 
larger set of solutions whose limiting cases are
\be
\begin{array}{lll}
^8B  & (1.29 - 2.1)\times10^{-8}eV^2                &  
(9.5\times10^{-8} - 1.3\times10^{-7})eV^2 \\
^7Be & (4.33\times10^{-9} - 1.32\times10^{-8})eV^2  &  \\
pp   & <5.6\times10^{-11}eV^2                       &  >6.8\times10^{-9}eV^2
\end{array}
\ee
for ${\mu}_{\nu}=7.2\times10^{-12}{\mu_B}$ and
\be
\begin{array}{lll}
^8B  & (3.30 - 7.32)\times10^{-9}eV^2              & 
(1.0 - 1.4)\times10^{-7}eV^2  \\
^7Be & (1.85\times10^{-9} - 1.51\times10^{-8})eV^2 & \\
pp   &    no~solution                              & >7.32\times10^{-9}eV^2\\
\end{array}
\ee
for ${\mu_{\nu}}=2.1\times10^{-11}{\mu_B}$, shown in fig.8. Hence, for the 
field (73), (75), the set of solutions is
\be
7.2\times10^{-12}{\mu_B}{\leq}{\mu_{\nu}}{\leq}2.1\times10^{-11}{\mu_B}~,~
7.3\times10^{-9}eV^2{\leq}{\Delta}^2m_{21}{\leq}1.3\times10^{-8}eV^2~.
\ee

The distinguishing feature between (74) and (75) is the same as between field
profiles (60), (62): two fall offs along the convective zone, one linear and 
the other with an upward facing concavity. The second type is clearly favoured
while the first is disfavoured. A downward facing concavity along the 
convective zone thus seems to be excluded. The type of exponential fall off
proposed years ago \cite{ALP1} where this downward concavity predominates,
namely (see fig.8),
\be
B(x)=\frac{B_0}{1+exp\frac{x-0.95}{0.01}}~~~,~~~0.7483{\leq}x{\leq}1
\ee
is also not possible. The main reason is that the moderate suppression 
required for $^8B$ neutrinos demands their resonances to be mainly located in
a zone where the field is substantially smaller than at the $^7Be$ resonance.
But then the $pp$ resonances are 'dragged' into a zone where the field is far 
too large. 

The main requirements for a suitable field, as from the analysis
of this section are therefore:

1) A very sharp rise across the bottom of the convective zone by at least
a factor of 6 - 7 over a distance not greater than approximately 9\% of the
solar radius. This sharp increase may be a discontinuity.

2) A decrease along the convective zone whose slope is at first strong and 
then reduces to become very moderate or even closer to constant  as one 
approaches the surface.

{\vspace{10mm}

{\large 5. Outlook and Conclusions}

{\vspace{6mm}

The solar neutrino data were used to obtain information or, possibly, 'derive'
the likely profile of the solar magnetic field on the basis of the existence
of a neutrino magnetic moment and the RSFP mechanism. The main point of the 
analysis relies on associating neutrino energy with  resonance location within
the Sun. The experimental data and general basic features of solar models
overwhelmingly favour an energy dependent neutrino suppression, the main 
uncertainty left being the amount of suppression of the highest energetic 
$^8B$ neutrinos. So neutrino suppression is likely to be correlated to 
resonance location within the Sun and thus to the profile of the field 
intensity, since a strong field at the resonance is responsible for a high 
transition rate. These are the underlying arguments for a possible tomography
of the solar magnetic field. 

On the basis of these arguments, the large suppression of the intermediate
energy neutrinos consisting mainly of the $^7Be$ line at $E=0.86$MeV, along
with the total or nearly total survival of the detected $pp$ neutrinos
$(0.236{\rm}MeV<E_{pp}<0.42{\rm}MeV)$ is a clear indication of a solar field
rising sharply along a relatively short distance. Such a topology can 
only be expected to be located around the bottom of the convective zone within
a layer of at most (9 - 10)\% of the solar radius. From the analysis of the 
previous section the field should rise by a factor of at least 6 - 7. All is 
consistent with this rise being extremely sudden, even in the form of a 
discontinuity, or to span most of that layer. If one believes in the RSFP 
mechanism, one may get an upper bound on the field (see eq.(46)) from the
inner edge of this zone down to approximately $0.59R_S$ to be 
$B_{max}{\simeq}(1.1~-~1.3)\times10^4G$, taking ${\mu_{\nu}}=10^{-11}{\mu_B}$.   
At the peak the field should be $B_{min}{\simeq}7\times10^4G$ for the same 
${\mu_{\nu}}$. Both these bounds are inversely proportional to the neutrino
magnetic moment. Deeper into the radiation zone the field is essentially 
unknown, although there is no ground to believe in other large spatial 
variations. 

As regards the convective zone, one expects a decreasing field with no sudden
variations, the analysis favouring a larger slope towards the bottom with a 
more moderate one towards the surface. The 'true' field profile thus appears 
to be consistent with any of the ones shown in figs.2 through 7, with fig.8 
excluded because of the shape of the field decrease along the convective zone.

In the original papers \cite{VVO} introducing the hypothetical anticorrelation
of active neutrino flux with solar activity as evidenced through the sunspots,
it was suggested that the sunspot activity effect could extend as deep as the
bottom of the convective zone. This is however by no means clear, hence the
extent of neutrino trajectory affected by this 11 year sunspot cycle is
unknown. The evaluation of the adiabaticity parameter $\gamma_C$ (eq.(25))
for all magnetic field distributions considered shows that only the one in
fig.5 (eq. (66) with $B=0$ for $x<0.71$) may generate moderately adiabatic
transitions for the Super-Kamio-kande neutrinos with energies ranging from
their 7MeV threshold up to 9MeV. Even this situation will only occur at the
upper edge of the solution, namely $\mu_{\nu}=1.3\times 10^{-10}\mu_B$, 
$\Delta^2m_{21}=1.6\times10^{-8}eV^2$, (eq. (72)), in clear conflict with the
astrophysical bounds on $\mu_{\nu}$ \cite{Abounds}. For all other cases the
Super-Kamio-kande neutrinos are expected to undergo non-adiabatic transitions
in the Sun, so no anticorrelation with sunspot activity is expected, even if
the effect extends deep down into the convective zone. For the Chlorine
experiment on the other hand, a larger fraction of $^8B$ neutrinos may be
anticorrelated, since the energy threshold is much lower, but the effect
may be totally unclear due to the dominant influence of the more energetic
neutrinos whose resonances are non-adiabatic. Therefore RSFP scenarios are
fully consistent with the absence of anticorrelation in the experiments so 
far \cite{W}.

As far as the approach used throughout this paper is concerned, a few words
must be added regarding the simplified treatment of the intermadiate energy 
neutrinos. An overall suppression factor of at least 96\% was considered for 
all of them, with a fixed energy $E=E_{^7Be}=0.86$MeV. They however span an 
energy range up to 1.7 MeV with this $^7Be$ line accounting for the largest 
part $(\simeq71\%)$. Nevertheless, a more elaborate procedure is hardly 
justified at the present stage of both data and theoretical models and should 
not change significantly the results derived here. In fact the suppression 
peak will necessarily be still assumed to occur at the 0.86MeV $^7Be$ line, 
with the other $^7Be$ line at 0.38 MeV, immersed in the $pp$ energy range, 
being naturally much less suppressed. The CNO continuum may also suffer this 
suppression peak around 0.86 MeV but, for the remainder, it need not, along 
with the pep line ($E=1.442 \rm{MeV}$), undergo a large suppression.

It should be stressed that all information that can be obtained from the RSFP
on the solar magnetic field only concerns its profile and not its absolute
magnitude because the order parameter of the analysis is the product ${\mu}B$.
The analysis and results refer only to the transverse component of the 
magnetic field along the direction of propagation
and nothing can be inferred as to its latitude dependence. They 
necessarily bear the uncertainties common to all solar physics models.

From the experimental side, more abundant statistics and precise data together
with new solar neutrino experiments are essential in order to improve our 
knowledge, along with independent laboratory searches for neutrino magnetic
moment effects. In has been shown in this paper that the necessary 
$\mu_{\nu}$ to provide a solution to the solar neutrino problem, thereby 
leading to information on the magnetic field, is within most of the relatively
stringent astrophysical bounds \cite{Abounds}. Although this lies almost 6
orders of magnitude above the electroweak standard model prediction, one 
should keep in mind that the standard model itself is full of fine tuning and
similarly large hierarchy problems affecting for instance fermion masses. 
Furthermore, there exist successful models \cite{BFZ} for large ${\mu_{\nu}}$
which do not conflict with the necessary smallness of neutrino mass.

{\vspace{10mm}}

{\large Acknowledgements}

{\vspace{5mm}}

The author wishes to express his gratitude to John Ralston for enlightening
discussions and to the Department of Physics and Astronomy of the University of
Kansas for hospitality and financial support. He is also grateful to Stanley
Brodsky and the SLAC Theory Group for providing a visitorship during which
part of this work was performed.

{\newpage}
\begin{center}
\begin{tabular}{cccc}                                     \hline\hline
Experiment      &  Ref.                    &  Data                         & 
Units\\ \hline
Homestake       &  \cite{Hom}              &  $2.54\pm0.14\pm0.14$         & 
SNU  \\
Kamiokande      &  \cite{Kamiokande}       &  $2.80\pm0.19\pm0.33$         & 
$10^6cm^{-2}s^{-1}$ \\ 
SAGE            &  \cite{SAGE}             &  $72\pm^{12}_{10}\pm^{5}_{7}$ & 
SNU  \\
Gallex          &  \cite{Gallex}           &  $76.2\pm6.5\pm5$             & 
SNU  \\
SuperKamiokande &  \cite{SuperK}           &  $2.51\pm^{0.14}_{0.13}$      &
$10^6cm^{-2}s^{-1}$ \\ \hline\hline
\end{tabular}
\end{center}
Table I. Neutrino event rates measured by solar neutrino experiments.

\vspace {10mm}

\begin{center}
\begin{tabular}{lcccccc}  \hline\hline
         & TCL\cite{TCL} & TCCCD\cite{TCCCD} & P94\cite{P94}    & 
RVCD96\cite{RVCD96} & FRANEC96\cite{FRANEC96} & BP95\cite{BP95} \\  \hline
Chlorine &               &                   &                  &
                    &                         &                 \\
pp       & 0.0           & 0.0               & 0.0              &
0.0                 & 0.0                     & 0.0             \\
pep      & 0.22          & 0.21              & 0.222            &
0.221               & 0.224                   & 0.224           \\
$^7$Be   & 1.10          & 0.995             & 1.24             &
1.15                & 1.08                    & 1.24            \\
$^{13}$N & 0.063         & 0.104             & 0.109            &       
0.095               & 0.0901                  & 0.105           \\
$^{15}$O & 0.21          & 0.37              & 0.379            &
0.327               & 0.306                   & 0.371           \\ 
$^8$B    & 4.63          & 4.06              & 7.19             &
7.03                & 5.73                    & 7.35            \\ \hline                
total    & 6.23          & 5.75              & 9.1              &
8.8                 & 7.4                     & 9.3             \\\hline\hline
Gallium  &               &                   &                  &
                    &                         &                 \\
pp       & 71.1          & 70.6              & 69.7             &
70.1                & 70.7                    & 69.7            \\
pep      & 2.99          & 2.795             & 2.99             &
2.97                & 3.01                    & 3.01            \\
$^7$Be   & 30.9          & 30.6              & 37.9             &
35.1                & 32.9                    & 37.7            \\
$^{13}$N & 2.36          & 3.87              & 3.95             & 
3.45                & 3.27                    & 3.82            \\
$^{15}$O & 0.21          & 6.5               & 6.46             &
5.58                & 5.22                    & 6.32            \\
$^8$B    & 10.77         & 9.31              & 15.7             &
15.4                & 12.5                    & 16.1            \\ \hline 
total    & 122.5         & 124               & 137              &
133                 & 128                     & 137             \\\hline\hline
Kam      &               &                   &                  &
                    &                         &                 \\
         & 4.4           & 3.8               & 6.48             &
6.33                & 5.16                    & 6.62            \\\hline\hline
\end{tabular}
\end{center}
Table II. Contributions of different sources of neutrinos to total capture 
rates in the Chlorine, Kamiokande and Gallium experiments from each standard 
solar model considered. All models include diffusion except for TCL \cite{TCL}
and TCCCD \cite{TCCCD}. Units are $10^6\rm{cm}^{-2}\rm{s}^{-1}$ for Kamiokande 
and SNU for Chlorine and Gallium. 

{\newpage}
Fig.1 Survival probability of $pp$ neutrinos with a magnetic moment 
$\mu_{\nu}=1.3\times10^{-10}\mu_{B}$ for the field distribution given by 
eq.(66) with $B=0$ in the radiative zone and core. The left end of the diagram 
corresponds to the resonance being close to the surface of the Sun or in the 
vacuum where the field is negligible, whereas the right end corresponds to a 
resonance in the interior (radiative zone and core) where the field also 
vanishes. \\

Fig.2 Field distribution given by eq.(50).  
Coordinate $x$ denotes the fraction of
the Sun's radius and the magnetic field is given in Gauss. The solution range
is given by eq.(57). \\

Fig.3 Field distribution given by eq.(62) ('option 1'). For the solution see
eq.(63). \\

Fig.4 Field distribution given by eq.(64) ('option 2'). For the solution see
eq.(65). \\

Fig.5 Field distribution given by eq.(66) with $B=0$ below the convective 
zone. The range of solutions is greatly extended relative to previous cases
(see eq.(72)) because the rise in the field at the bottom of the 
convective zone is both very large and sudden so that inequality (45) is 
largely satisfied. Also from the bottom of the convective zone to the solar 
surface the field has the 'adequate' shape that can account for a large 
range in terms of the $^8B$ solution. \\

Fig.6 Field distribution given by eqs.(73), (74). The range of solutions is
very limited (see eq.(78)).\\

Fig.7 Field distribution given by eqs.(73), (75). Although the magnitude of
the field is no larger than in the previous case, the solution is greatly
extended (compare eq. (78) with eq.(81)) owing to the difference in shapes. \\

Fig.8 Field distribution given by eq.(82). Although this distribution has the
correct shape across the bottom of the convective zone with a large and sharp
increase, it fails to provide a solution, owing to its shape along the
convective zone.

\newpage

\begin{figure}
\begin{picture}(18,17)
\put(0,1){\psfig{figure=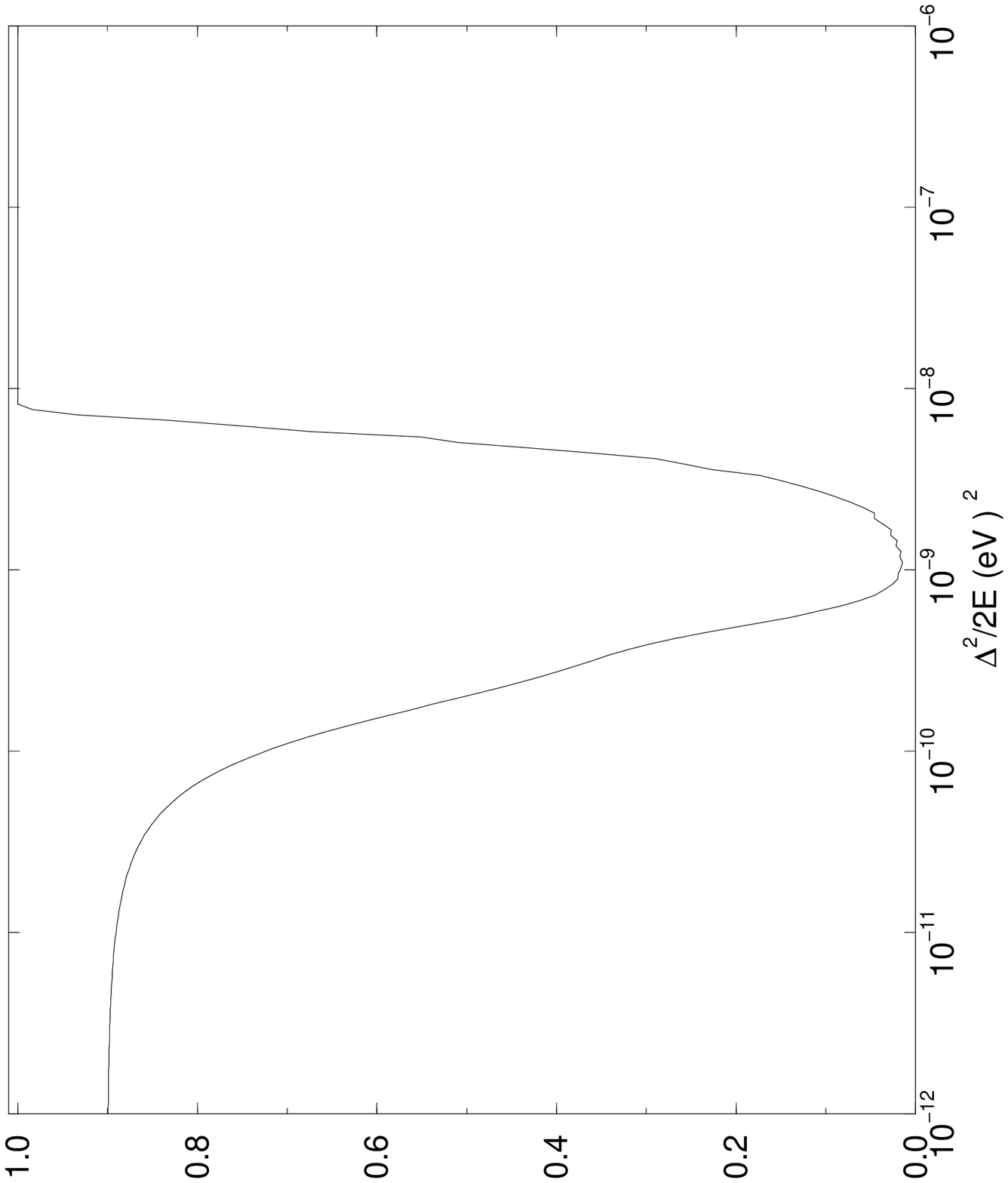,width=17cm}}
\end{picture}
\caption{}
\end{figure}

\begin{figure}
\begin{picture}(18,20)
\put(1,2){\psfig{figure=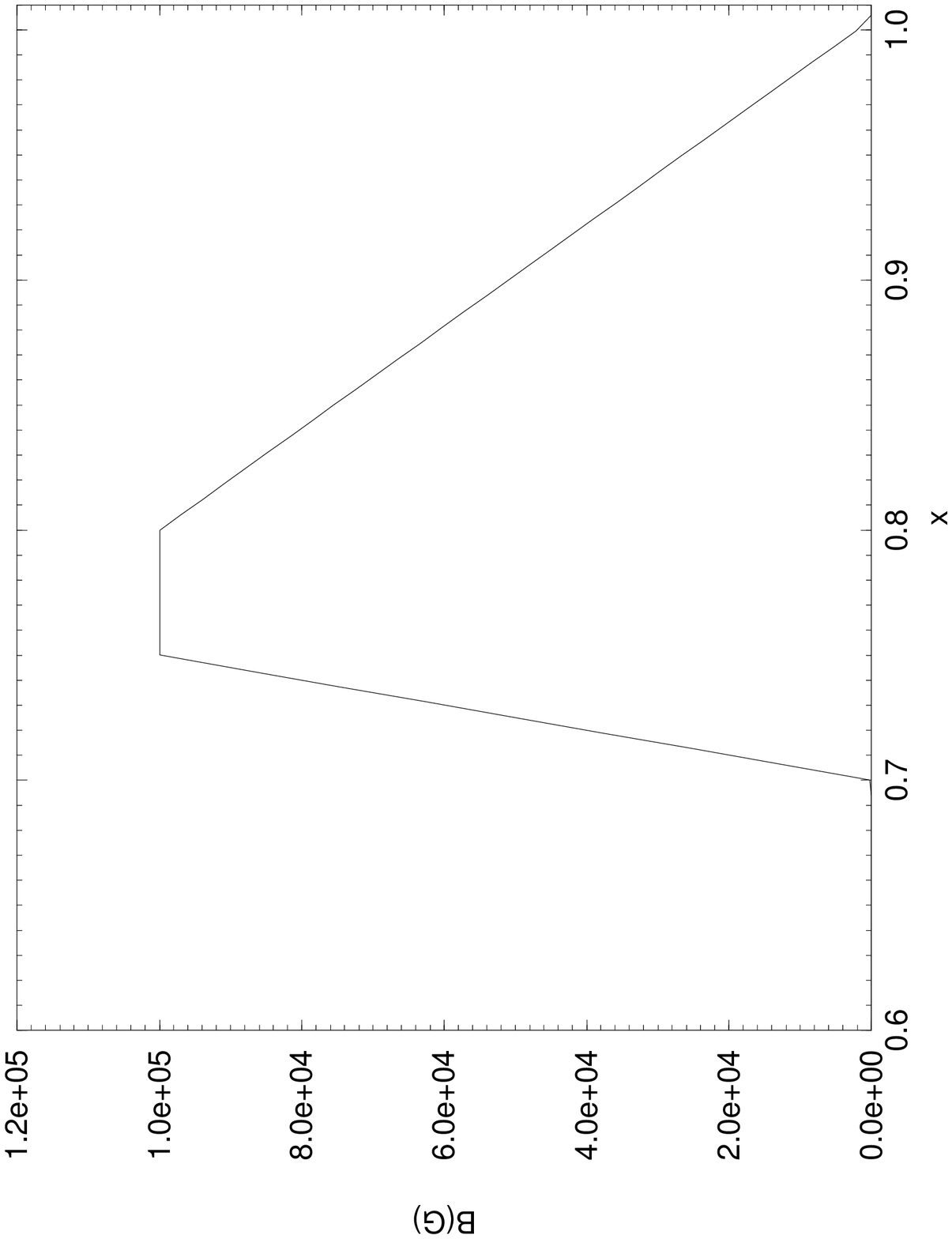,width=15cm}}
\end{picture}
\caption{}
\end{figure}

\begin{figure}
\begin{picture}(18,20)
\put(1,2){\psfig{figure=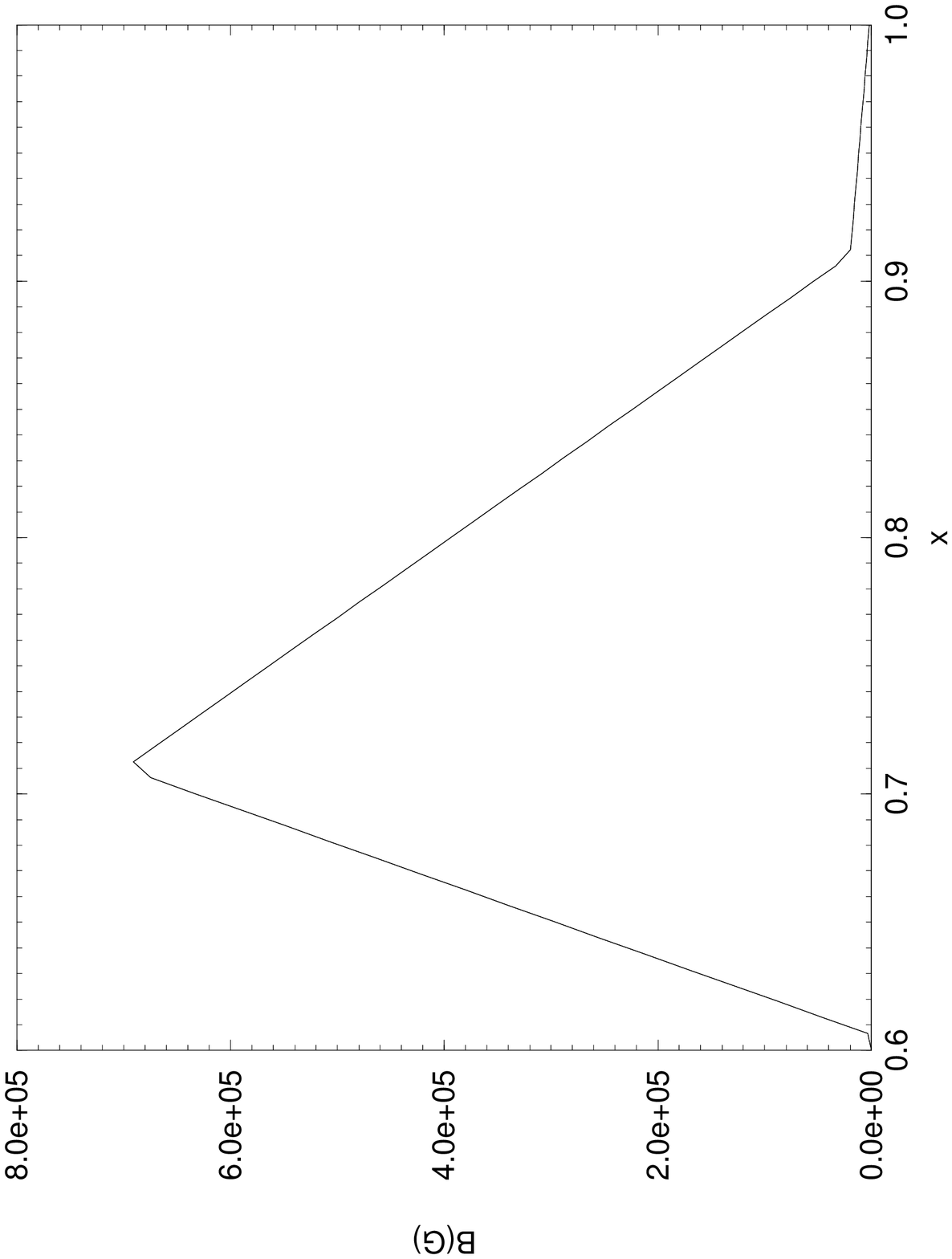,width=15cm}}
\end{picture}
\caption{}
\end{figure}

\begin{figure}
\begin{picture}(18,20)
\put(1,2){\psfig{figure=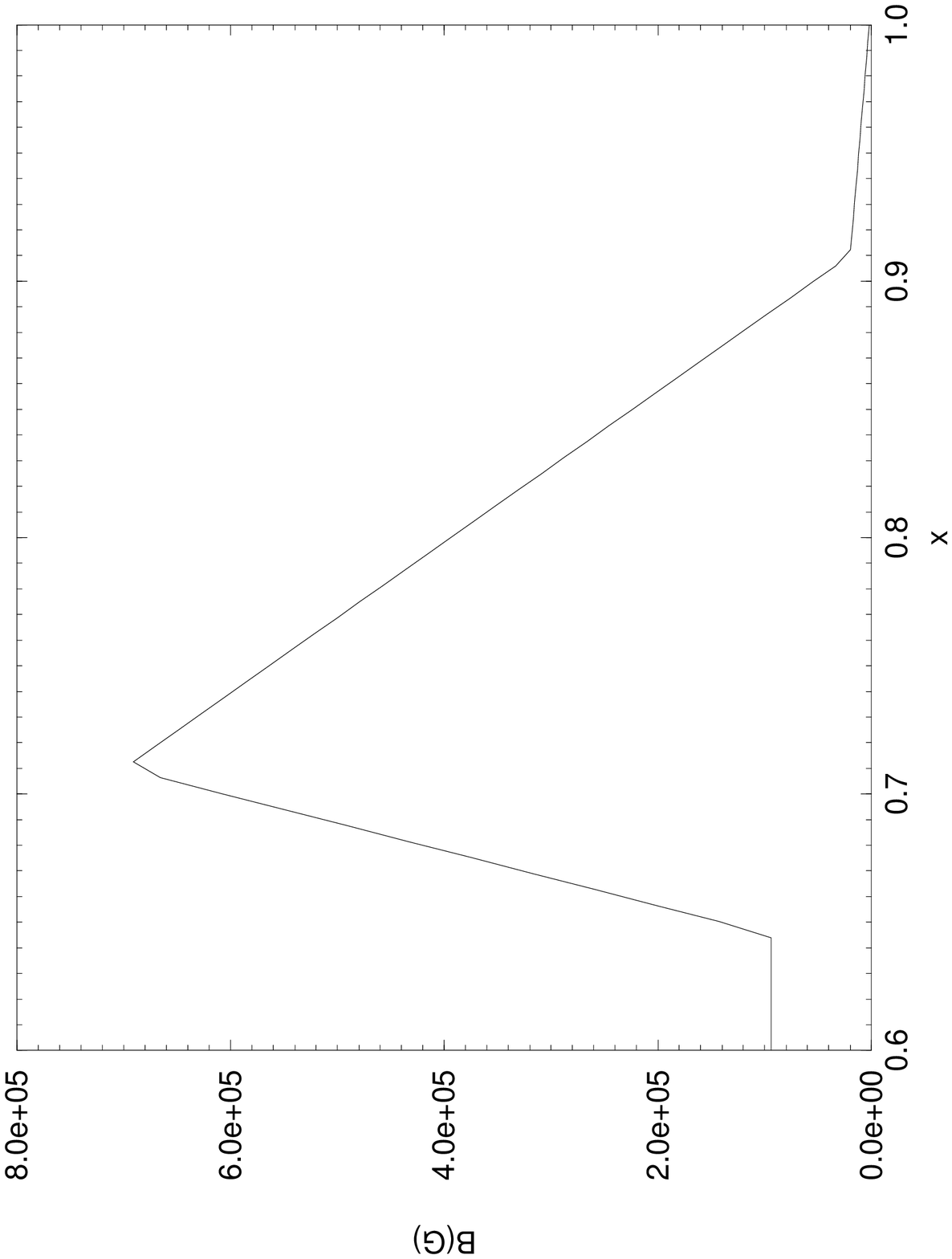,width=15cm}}
\end{picture}
\caption{}
\end{figure}

\begin{figure}
\begin{picture}(18,20)
\put(1,2){\psfig{figure=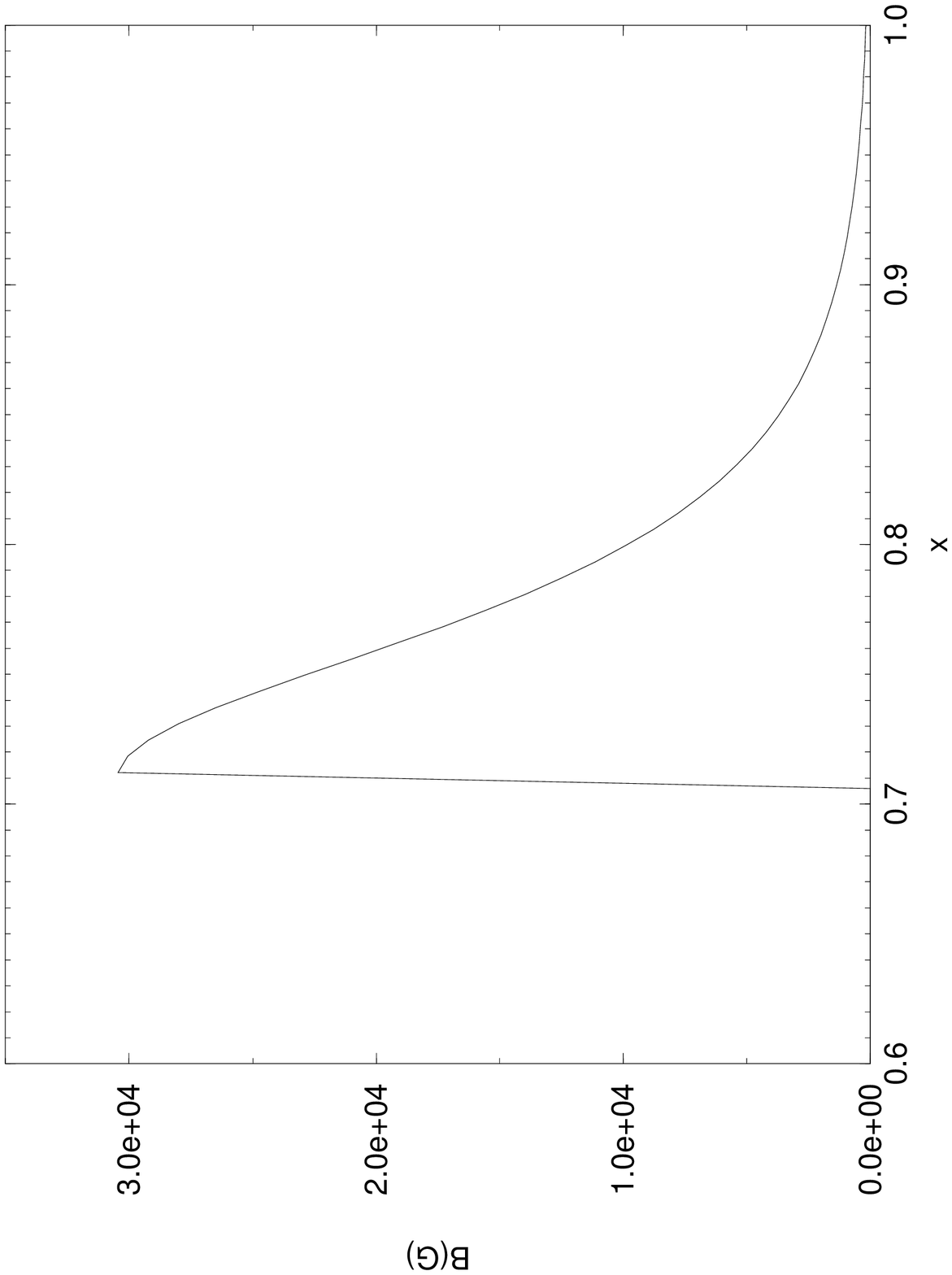,width=15cm}}
\end{picture}
\caption{}
\end{figure}

\begin{figure}
\begin{picture}(18,20)
\put(1,1){\psfig{figure=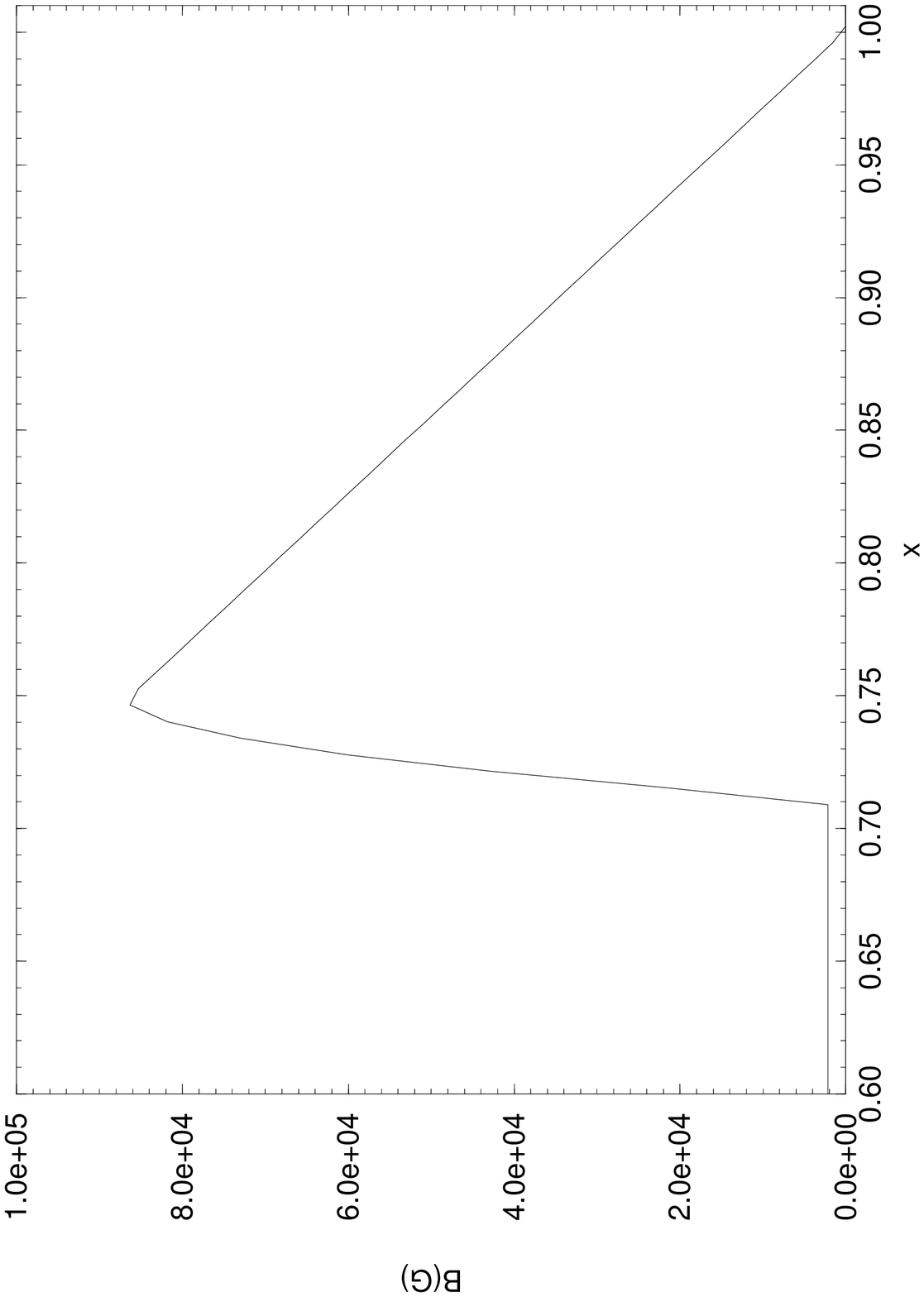,width=15cm}}
\end{picture}
\caption{}
\end{figure}

\begin{figure}
\begin{picture}(18,20)
\put(1,2){\psfig{figure=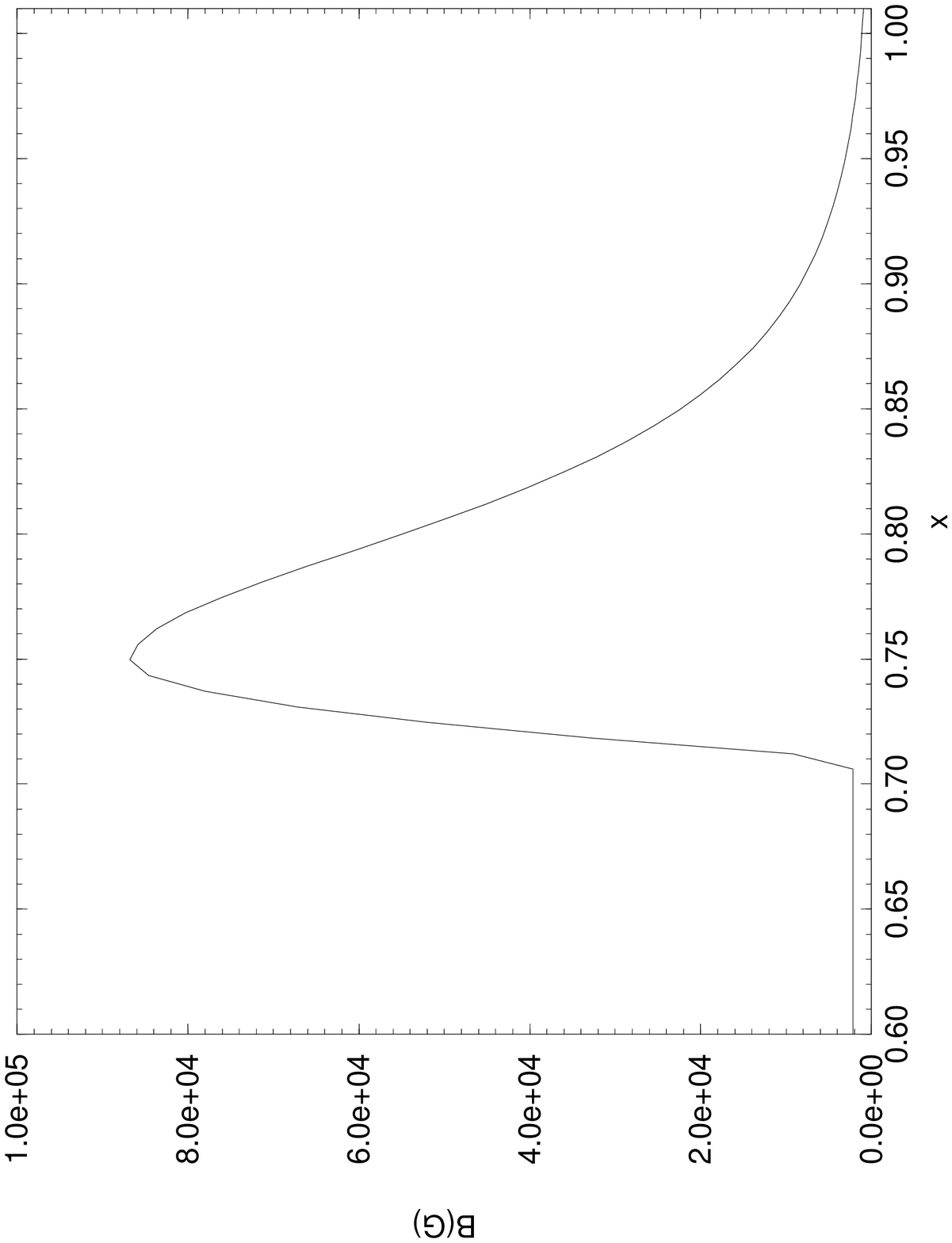,width=15cm}}
\end{picture}
\caption{}
\end{figure}

\begin{figure}
\begin{picture}(18,20)
\put(1,1){\psfig{figure=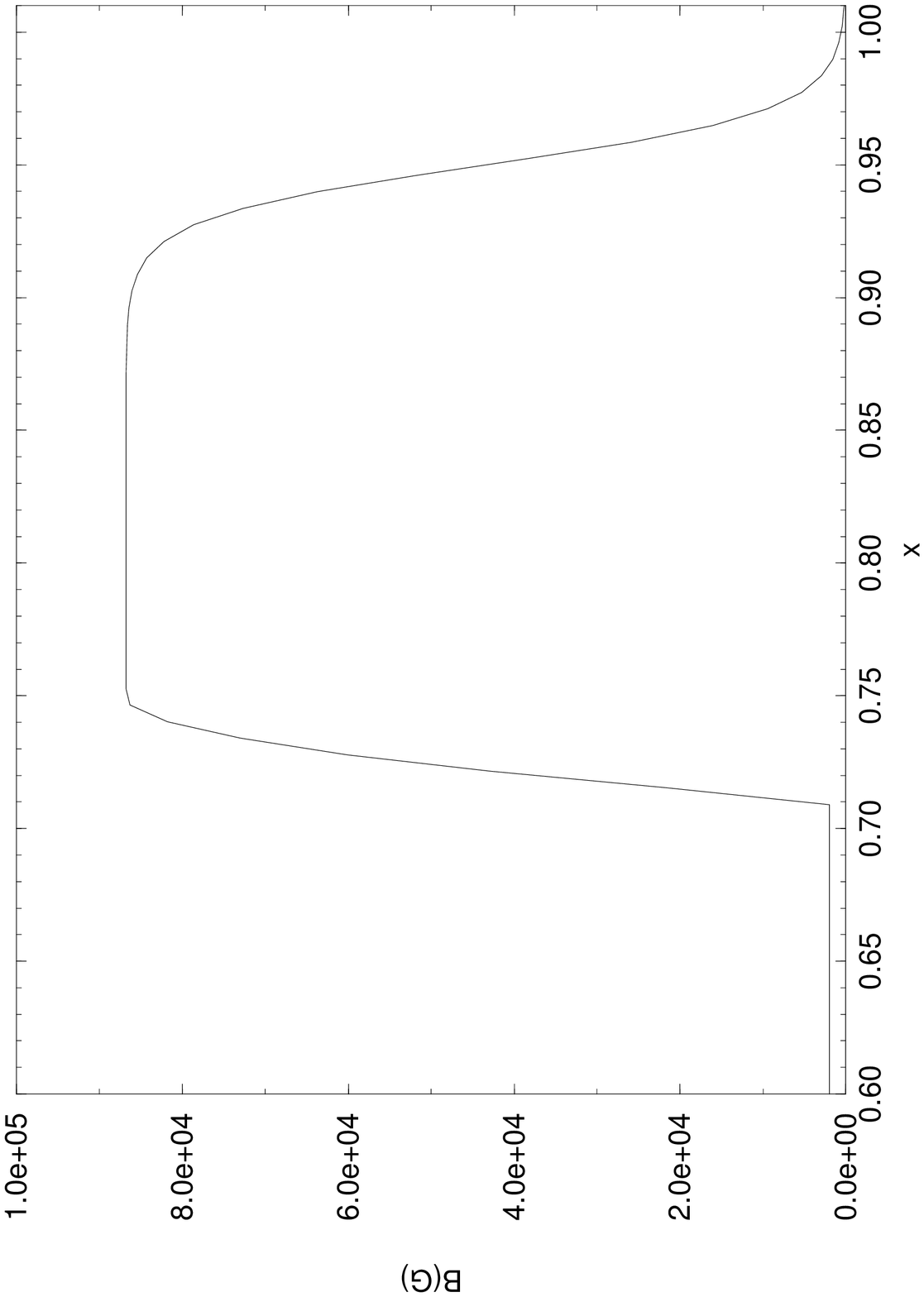,width=15cm}}
\end{picture}
\caption{}
\end{figure}

\end{document}